\documentclass[pra,showpacs,superscriptaddress]{revtex4-1}
\usepackage[psamsfonts]{amssymb}
\usepackage{graphicx}
\usepackage{amsmath}
\usepackage{mathrsfs}
\usepackage{subcaption}
\newcommand{\bra}[1]{\langle #1 \vert}
\newcommand{\ave}[1]{\langle #1 \rangle}
\newcommand{\ket}[1]{\vert #1 \rangle}
\newcommand{\la}{\langle}
\newcommand{\ra}{\rangle}
\newcommand{\commentold}[1]{}
\DeclareMathSymbol{:}{\mathpunct}{operators}{"3A}
\begin{document}
%%%%%%%%%%%%%%%%%%%%%%%%%%%%%%%%%%%%%%%%%%%%%%%%%%%%%%%%%%%%%%%%%%%%%%%%%%%%%%%%%%%%%%%%%%%%%%%%%%%%%%%%%%%%%%%%%%%%%%%%%%%%%%%%%%%%%%%%%%%%%%%%%%%%%%%%%%%%%%%%%%%%%%%%%%%%%%%%%
\title{Quantum thermodynamics of a trapped two-level atom in an external light field }
\author{A. Moradian}
\affiliation{Department of Physics, University of Kurdistan, P.O.Box 66177-15175, Sanandaj, Iran}
\author{F. Kheirandish}
\email{kheirandish@uok.ac.ir}
\affiliation{Department of Physics, University of Kurdistan, P.O.Box 66177-15175, Sanandaj, Iran}
\affiliation{Department of Physics, University of Garmian, Kalar, KRG, Iraq}
\date{\today}% It is always \today, today,
\begin{abstract}
Quantum thermodynamics of a trapped two-level atom under the influence of a controlled light field is investigated. The population dynamics and decoherency function are obtained and discussed. The characteristic functions, work distribution functions and Helmholtz free energies are calculated and the consistency with the Jarzynski theorem is verified.
\end{abstract}
%%%%%%%%%%%%%%%%%%%%%%%%%%%%%%%%%%%%%%%%%%%%%%%%%%%%%%%%%%%%%%%%%%%%%%%%%%%%%%%%%%%%%%%%%%%%%%%%%%%%%%%%%%%%%%%%%%%%%%%%%%%%%%%%%%%%%%%%%%%%%%%%%%%%%%%%%%%%%%%%%%%%%%%%%%%%%%%%%%
%\pacs{42.50.Ct, 42.50.Nn, 03.70.+k, 78.20.Ci, 78.67.Pt}

%\keywords{Quantum thermodynamics; Decoherency; Characteristic function; Work distribution function}

\date{\today}
\maketitle
%
%%%%%%%%%%%%%%%%%%%%%%%%%%%%%%%%%%%%%%%%%%%%%%%%%%%%%%%%%%%%%%%%%%%%%%%%%%%%%%%%%%%%%%%%%%%%%%%%%%%%%%%%%%%%%%%%%%%%%%%%%%%%%%%%%%%%%%%%%%%%%%%%%%%%%%%%%%%%%%%%%%%%%%%%%%%
\section{Introduction}\label{Introduction}
%%%%%%%%%%%%%%%%%%%%%%%%%%%%%%%%%%%%%%%%%%%%%%%%%%%%%%%%%%%%%%%%%%%%%%%%%%%%%%%%%%%%%%%%%%%%%%%%%%%%%%%%%%%%%%%%%%%%%%%%%%%%%%%%%%%%%%%%%%%%%%%%%%%%%%%%%%%%%%%%%%%%%%%%%%%
Thermodynamics of small systems in the realm of quantum mechanics deals with quantum fluctuations of the same order of magnitude as expectation values \cite{tasaki2000jarzynski,Roeck2004quantum,talkner2007tasaki,deffner2008nonequilibrium}. Therefore, the standard approach of thermodynamics is not applicable in the quantum domain for small systems and a precise knowledge of fluctuation-theorems is unavoidable. Fluctuation theorems are a bridge between non-equilibrium fluctuations and thermal equilibrium states of a small system. They can also describe the nonlinear response of a system to external forces. An extensive work has been done by researchers to find the correct behaviour of quantum dynamics of small nonequilibrium systems. Due to the important rule played by the small quantum systems interacting with their environment, the quantum thermodynamics of such systems is an active subject nowadays. The important distribution functions like heat and work distribution and physical quantities like irreversibility of work, kinds of entropy and Helmholtz free energy have been investigated \cite{blickle2006thermodynamics,harbola2006quantum,esposito2006fluctuation,kheirandish2020many,campisi2011erratum,
talkner2008microcanonical,Roeck2004quantum,deffner2008nonequilibrium,crooks1999entropy,jarzynski2004nonequilibrium,talkner2007tasaki,jarzynski2011equalities,evans1993probability}.

An important quantity in the context of quantum thermodynamics is the amount of work that can be extracted or done on a certain quantum system out of equilibrium. But there is not an observable corresponding to work \cite{talknerLutzHangi2007}. Therefore, some schemes have been introduced to measure work and the most established method is the two-point measurement scheme \cite{EspositoHarbolaMukamel2009}. Here we apply the two-point measurement scheme to a trapped two-level atom under the influence of an external laser field.
Trapped two-level atoms have been investigated in a wide range of problems like atom-field entanglement \cite{cereceda2000quantum,schrodinger1935gegenwartige}, quantum computations, squeezing \cite{walls1983squeezed,slusher1985observation,vahlbruch2008,kimble1977photon}, quantum correlations \cite{gerber2009intensity,rempe1987observation}, coherency, revivals and collapses \cite{rempe1987observation,lipfert2018time,kimble1977photon}, anti-bunching of photons \cite{kimble1977photon}, Schr\"{o}dinger cat states \cite{brune1992manipulation,slosser1989harmonic,guo1996generation}, Fock states \cite{slosser1989harmonic,weidinger1999trapping,brattke2001generation}, inversion of the population of the states and Rabi oscillations \cite{rabi1936process,esteve2004quantum}. The Jayne's-Cummings model has been applied in a wide range of applications like, cooper pair box circuits \cite{wallraff2004strong}, superconducting flux qubits \cite{chiorescu2004coherent},
Josephson junctions \cite{hatakenaka1996josephson,sornborger2004superconducting}, quantum dots, and couplings of qubits to cavity modes
\cite{meier2004spin,basset2013single,kasprzak2010up,Freitas2017Josephson}.

The layout of the present work is as follows: In Sec. \ref{Atom-light-classical}, interaction of a two-level atom with a classical light field controlled by a switching function is investigated analytically and numerically. The probability of level-occupation, decoherency function, work distribution function and Helmholtz free energy are obtained and discussed. In Sec. \ref{generalisation to vibrating atom}, the approach is generalised to a more realistic case of a trapped two-level atom vibrating in a harmonic potential under the influence of a classical light field. Finally, we conclude in Sec. \ref{conclusion}.
%%%%%%%%%%%%%%%%%%%%%%%%%%%%%%%%%%%%%%%%%%%%%%%%%%%%%%%%%%%%%%%%%%%%%%%%%%%%%%%%%%%%%%%%%%%%%%%%%%%%%%%%%%%%%%%%%%%%%%%%%%%%%%%%%%%%%%%%%%%%%%%%%%%%%%%%%%%%%%%%%%
\section{Interaction of a two-level atom with a classical light field \label{Atom-light-classical}}
\begin{figure}[h]
	\centering
			\includegraphics[width=.5\textwidth]{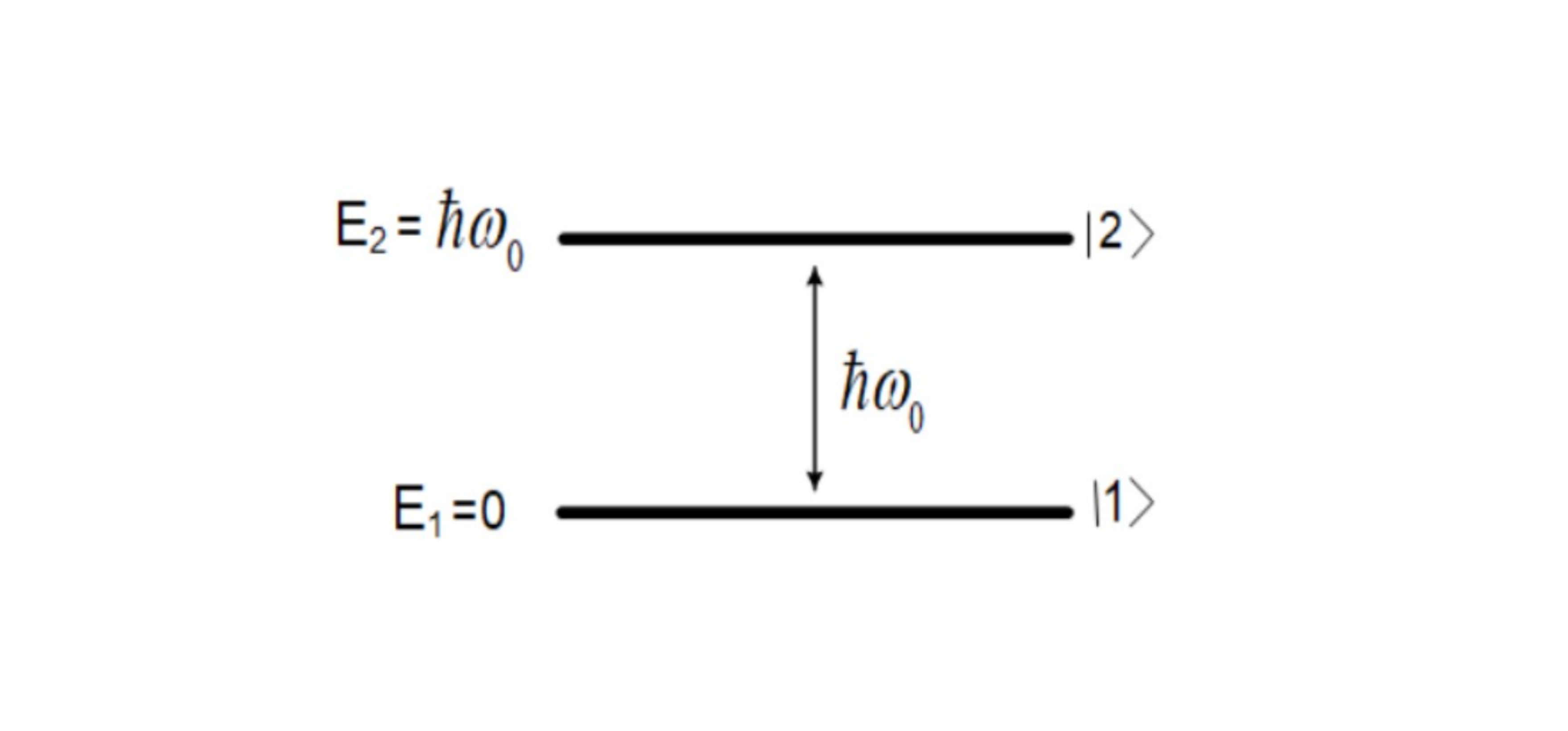}					
\caption{A two-level atom}
	\label{2level atom.Fig}
\end{figure}

%%%%%%%%%%%%%%%%%%%%%%%%%%%%%%%%%%%%%%%%%%%%%%%%%%%%%%%%%%%%%%%%%%%%%%%%%%%%%%%%%%%%%%%%%%%%%%%%%%%%%%%%%%%%%%%%%%%%%%%%%%%%%%%%%%%%%%%%%%%%%%%%%%%%%%%%%%%%%%%%%%
Let $\mathbf{E}(t)=\boldsymbol{\epsilon}\, E_0\,\cos\omega_L t$ be the electrical component of a classical light field with frequency $\omega_L$, amplitude $E_0$, and polarization unit vector $\boldsymbol{\epsilon}$. Here we use the dipole approximation that is the wavelength of the applied classical light is much larger than the Bohr's atomic radius. Therefore, the electric field can be considered at the center of mass of the atom. The interaction of a two-level atom Fig.(\ref{2level atom.Fig}) interacting with a classical light field can be described by the Hamiltonian
\begin{align}
	\label{H}
	H=\hbar\,\omega_0\, \sigma^{+}\sigma^{-} + g(t)\, H_{int},
\end{align}
where the first term is the Hamiltonian of the two-level atom and the second term is the interaction Hamiltonian $H_{int}=-\mathbf{d}\cdot \mathbf{E}(t)$. The interaction term originates from the coupling between the electrical component of the light field and the atomic dipole $\mathbf{d}=\mathbf{d}_{12} \sigma^{-}+\mathbf{d}^*_{12}\sigma^{+}$ where $\mathbf{d}_{12}=\langle 1|\mathbf{d}|2\rangle$ and $\sigma^{+}=  |2\rangle \langle1|$ and $\sigma^{-}=  |1\rangle  \langle2|$ are the corresponding atomic ladder operators. The time-dependent function $g(t)$ in Eq. (\ref{H}) is a switching function that controls the interaction between the atom and the classical light.
%
%%%%%%%%%%%%%%%%%%%%%%%%%%%%%%%%%%%%%%%%%%%%%%%%%%%%%%%%%%%%%%%%%%%%%%%%%
\subsection{Hamiltonian of  the system in the rotating frame}
%%%%%%%%%%%%%%%%%%%%%%%%%%%%%%%%%%%%%%%%%%%%%%%%%%%%%%%%%%%%%%%%%%%%%%%%%
To find the time-evolution operator corresponding to Eq. (\ref{H}) we can transform the Hamiltonian by a unitary operator $R=\exp(-i\omega_L t\,\sigma^+\sigma^-)$ to a new Hamiltonian $\tilde{H}$ in the rotating frame as
\begin{equation} \label{Htilde}
	\tilde{H}=-i \hbar R^\dagger \dot{R}+ R^\dagger H R.
\end{equation}
By inserting the operator $R$ into Eq. (\ref{Htilde}) one easily finds

\begin{align}\label{Htilde2}
\tilde{H}=\hbar\,\Delta\,\sigma^+ \sigma^--\hbar\,\Omega\, g(t)\left(e^{-i \theta} \sigma^++h.c.\right)-\hbar\,\Omega\, g(t)\left(e^{-i \theta} \sigma^+ e^{2 i \omega_L t}+h.c.
\right),
\end{align}
where $\Delta=\omega_0 -\omega_L$ is the detuning frequency. Here for simplicity we have defined $E_0 \,\boldsymbol{\epsilon}\cdot \mathbf{d}_{12}/2 \hbar=\Omega\,e^{i\theta}$, whose modulus is $\Omega$ with a phase $\theta$. The parameter $\hbar \Omega $ is a strength coupling proportional to the projection of the electrical field in the direction of the atomic dipole. From an analytical point of view, due to the rapid oscillations, the RWA(rotating wave approximation) can be applied ($\omega_L\gg0$) and the last term in Eq. (\ref{Htilde2}) can be ignored. In the following we also keep the last term in Eq. (\ref{Htilde2}) and find a much more precise result using a numerical approach. The switching function $g(t)$ is an arbitrary function up to the initial condition $g(0)=0$ and it is implicitly assumed that the switching function is zero after a time $\tau$ ($g(t>\tau)=0$) indicating the duration of interaction. Here we assume that $g(t)$ is a periodic function $g(t)=g(t+p)$ with period $p$ for ($t<\tau$) and $\tau$ is an integer multiplication of $p$. In the following, to find explicit numerical results we have set $p=2$ and assumed
\begin{equation} \label{gfunction}
g(t) =
\begin{cases}
	1, & \text{for } 0 \leq t\leq 1,
  \\
	0, & \text{for } 1 < t\leq 2.
\end{cases}
\end{equation}
%%%%%%%%%%%%%%%%%%%%%%%%%%%%%%%%%%%%%%%%%%%%%%%%%%%%%%%%%%%%%%%%%%%%%%%%%%%%%%%%%%%%%%%%%%%%%%%%%%%%%%%%%%%%%%%%%%%%%%%%%%%%%%%%%%%%%%%%%%%%%%%%%%%%%%%%%%	
\subsection{The unitary evolution: the on-resonance case}
In the on-resonance case $\Delta=0$ and by making use of RWA, we will find that the rotating frame Hamiltonian $ \tilde{H}$ commutes at different times $ [\tilde{H}(t),\tilde{H}(t^{'})]=0$, therefore, in the rotating frame, the time-evolution operator can be obtained simply as
\begin{align}\label{utilde}
\tilde{U}(t,0)={\mathrm{e}}^{-\dfrac{i}{\hbar}\int\limits_0^t\! \tilde{H}(t')\, \mathrm{d}t'}.
\end{align}
The explicit expression of the evolution operator in matrix representation is
\begin{align} \label{Utilde}
\tilde{U}(t,0)=\begin{bmatrix}
\cos(G\,\Omega) & i e^{i \theta} \sin(G\,\Omega)
\vspace{6mm}
\\i e^{-i \theta} \sin(G\,\Omega) & \cos(G\,\Omega)
\end{bmatrix},
\end{align}
where $G(t)=\int\limits_{0}^{t} g(t^{'}) dt^{'}$.

To find the time-evolution operator in the lab frame, we use

 \begin{align}
 \label{Unitary evolution RWA}
U(t,0)=R(t,0) \tilde{U}(t,0),
\end{align}
 where
$ \label{eq.tilde u for atom in classiac laser}
R(t,0)=\begin{bmatrix}
1&0
\vspace{6mm}
\\0 & e^{-i\omega_L t}
\end{bmatrix}.
$
For the non-resonance case see App. (\ref{appendixA}).
%%%%%%%%%%%%%%%%%%%%%%%%%%%%%%%%%%%%%%%%%%%%%%%%%%%%%%%%%%%%%%%%%%%%%%%%%%%%%%%%%%%%%%%%%%%%%%%%%
\subsubsection{The probabilities of level-occupation}
%%%%%%%%%%%%%%%%%%%%%%%%%%%%%%%%%%%%%%%%%%%%%%%%%%%%%%%%%%%%%%%%%%%%%%%%%%%%%%%%%%%%%%%%%%%%%%%%%
Let the initial density matrix of the two-level atom be given in the general form
\begin{align}\label{initial ro}
  \rho(0)=\left(
            \begin{array}{cc}
              A & \xi^* \\
              \xi & 1-A \\
            \end{array}
          \right),
\end{align}
then the evolved density matrix at time $t$ is given by
\begin{equation}\label{evolved ro}
  \rho(t)=U(t,0)\,\rho(0)\,U^\dag (t,0).
\end{equation}
By inserting the matrix form of $U(t,0)=R(t,0)\,\tilde{U}(t,0)$ and $\rho(0)$ into Eq. (\ref{evolved ro}), we easily find
\begin{eqnarray}\label{ro components}
  \rho_{11} (t) &=& A\,\cos^2 (G \Omega)+(1-A)\,\sin^2 (G \Omega)-\sin(2 G \Omega)\,\mbox{Im}[\xi\,e^{i\theta}],\nonumber\\
  \rho_{12} (t) &=& \xi\,e^{2 i\theta}\,\sin^2 (G \Omega)+\xi^* \,e^{i\omega_L t}\,\cos^2 (G \Omega)+\frac{i\sin(2 G \Omega)}{2}\big[(1-A)\,e^{i(\theta+\omega_L t)}-A\,e^{i\theta}\big], \nonumber\\
  \rho_{21} (t) &=& \rho^*_{12} (t), \nonumber\\
  \rho_{22} (t) &=& 1- \rho_{11} (t),
\end{eqnarray}
where $\mbox{Im}[z]$ returns the imaginary part of $z$. For getting rid  of superfluous notation the arguments of $G$ has been dropped.
For the special case $A=1$, i.e. $\rho(0)=|1\rangle\langle 1|$, and $\xi=0$, we have
\begin{align}
   \label{eq. rho_0 for atom in classiac laser}
\rho(t)=\begin{bmatrix}
\cos^2(G \Omega)&-\frac{1}{2} i\,e^{i \theta} \sin (2 G \Omega )
\vspace{6mm}
\\\frac{1}{2} i\,e^{-i\theta}\sin (2 G \Omega )  &\sin^2(G \Omega)
\end{bmatrix}.
\end{align}

\begin{figure}[!tp]
	\centering
	\begin{subfigure}[b]{0.48\textwidth}
		\centering
		\includegraphics[width=\textwidth]{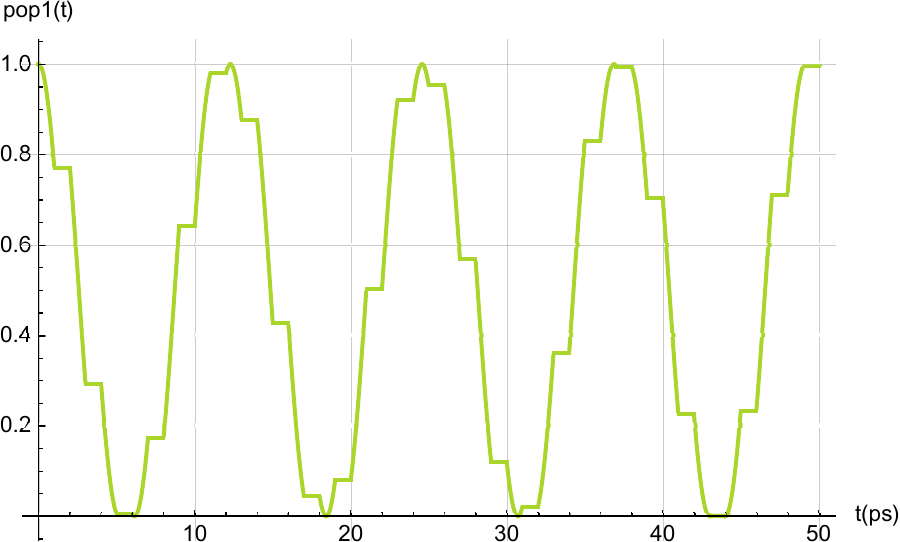}
		\caption{If we assume RWA is valid}
		\label{Fig.pop1}
	\end{subfigure}
	\hfill
	\begin{subfigure}[b]{0.48\textwidth}
		\centering
		\includegraphics[width=\textwidth]{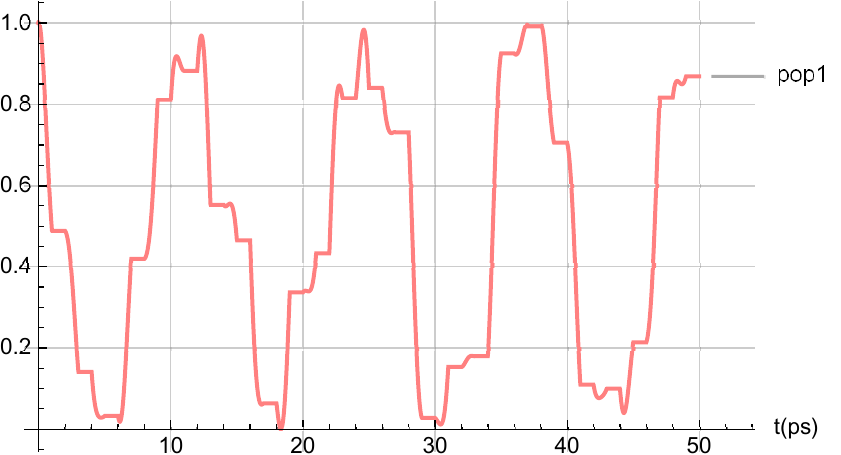}
		\caption{Without assuming RWA and keeping all terms in Eq. (\ref{Htilde2})}
		\label{Figpop1ND}
	\end{subfigure}
	\hfill
	
	\caption{The ground state occupation probability in time (picosecond) when the atom is initially prepared in the ground state. The parameters are chosen as $\Omega=.5$ \mbox{teraHz}, $\omega=1$ \mbox{teraHz}, $\xi=0 $, and $\theta=0$. }
		\label{popi and decoh}
\end{figure}
The probability of finding the system in its ground state is
 \begin{align}
 \label{Eq.pop RWA}
 pop_1 (t)= {|\bra{1}\rho(t)\ket{1}|} ^2.
\end{align}
By inserting Eq.(\ref{eq. rho_0 for atom in classiac laser}) into Eq.(\ref{Eq.pop RWA}) the function $pop_1 (t)$ has been depicted in Fig.(\ref{Fig.pop1}) where initially the system is prepared in ground state $\rho(0)=|1\rangle\langle 1|$. From Fig.(\ref{Fig.pop1}) we see that the ground state starts to evacuate during the successive pulses up to the time fulfilling $G \Omega =\frac{\pi}{2}$, then it rests while the pulse applying on the atom have been off. Using a numerical approach and without RWA, we have depicted $pop_1 (t)$ again in Fig.(\ref{Figpop1ND}) which shows almost the same behaviour but with different corresponding values.
\subsubsection{The dynamics of decoherency}
\begin{figure}[!tp]
	\centering
	\begin{subfigure}[b]{0.48\textwidth}
		\centering
		\includegraphics[width=\textwidth]{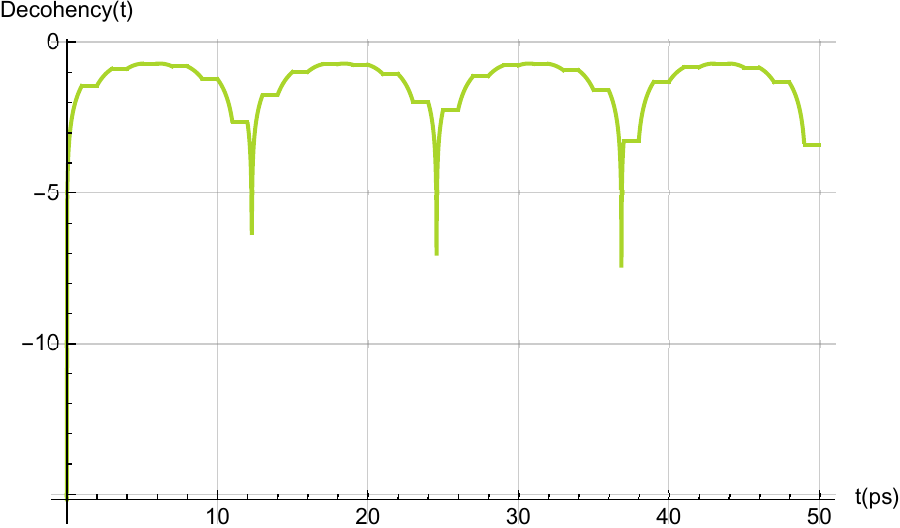}
		\caption{If we assume RWA is valid}
		\label{Fig.decoh}
	\end{subfigure}
	\hfill
	\begin{subfigure}[b]{0.48\textwidth}
		\centering
		\includegraphics[width=\textwidth]{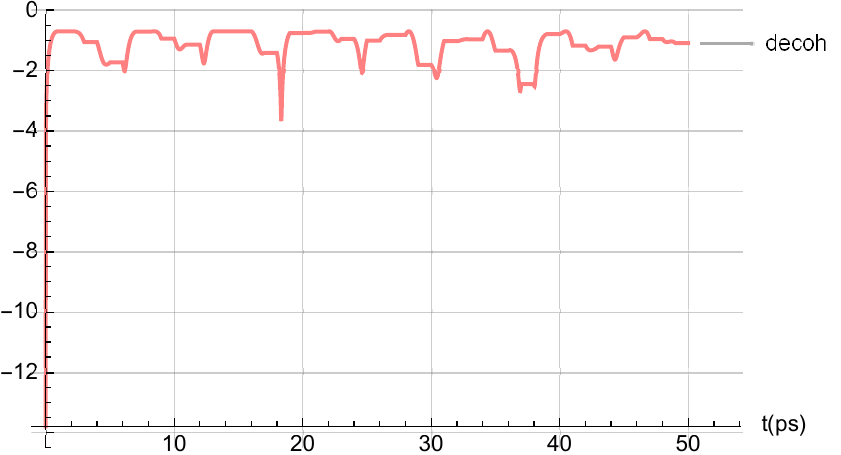}
		\caption{Without assuming RWA and keeping all terms in Eq. (\ref{Htilde2})}
		\label{Fig.decohND}
	\end{subfigure}
	\hfill
	
	\caption{The decoherency function versus time (in picosecond) when the atom is initially prepared in the ground state. The parameters are chosen as $\Omega=.5$ \mbox{teraHz}, $\omega=1$ \mbox{teraHz}, $\xi=0 $, and $\theta=0$.}
	\label{popi and decoh}
\end{figure}

The dynamics of the decoherency can be obtained from the definition
\begin{align}
\gamma (t)&=\log \left(\left\| \rho _{12}\right\| \right)
\nonumber
\\
&=-\log (2)+
\log (\left\| \sin ( G\Omega  )\right\|),
\end{align}
and is depicted in Fig.(\ref{Fig.decoh}) assuming RWA. It is seen that when the system evolves in time the states initiate to mix so the decoherency increases. It is evident that the decoherency plunges fast as it is evacuated. In Fig.(\ref{Fig.decohND}), the decoherency is depicted without taking into account RWA using a numerical procedure. In the right figure, the fluctuation domain of the decoherency is lesser compared to the left one.
\subsection{The characteristic function of work: on-resonance case}
The function $ ch(\nu,t)) $ which is the Fourier transform of the work distribution function is defined by
\begin{eqnarray} \label{definition of charastric function for a two level atom}
	ch(\nu,t))=Tr[e^{-i \nu H_0(0)} U^\dagger(t,0)e^{i \nu H(t)}U(t,0) \tilde{\rho}(0)],
\end{eqnarray}
where $U(t,0) $ is the time-evolution operator in the Schr\"{o}dinger picture from the initial time $t=0$ to $t$ and $\tilde{\rho}(0) $ is the diagonal part of the initial density matrix. By inserting Eq. (\ref{Unitary evolution RWA}) and Eq. (\ref{initial ro}) into Eq. (\ref{definition of charastric function for a two level atom}), we will find \cite{kheirandish2020many,talkner2008microcanonical,lipfert2018time}
\begin{align}
ch{(\nu,t)}&=\boldsymbol{i}\frac{ (1-2 A) \omega_0   \cos (2 G \Omega ) \sin \left(\frac{\alpha  \nu  \hbar }{2}\right) \cos \left(\frac{ \omega_0  \nu  \hbar }{2}\right)}{\alpha }+\frac{\omega_0   \cos (2 G \Omega ) \sin \left(\frac{\alpha  \nu  \hbar }{2}\right) \sin \left(\frac{ \omega_0 \nu   \hbar }{2}\right)}{\alpha }
\nonumber\\
+&\cos (\frac{\alpha  \nu  \hbar }{2}) \cos (\frac{\omega_0\nu    \hbar }{2})-\boldsymbol{i} (1-2 A) \cos \left(\frac{\alpha  \nu  \hbar }{2}\right) \sin \left(\frac{\omega_0\nu    \hbar }{2}\right)
\label{def of charactristic RWA},
\end{align}
where $\alpha=\sqrt{{\omega_0}^2+4 {G}^2 \Omega^2}$ denotes an effective frequency. The averaged work can be obtained from the characteristic function as
\begin{align} \label{eq. the first moment of atom in ground state ,laser is on}
\ave{W}= \dfrac{\partial ch(\nu,t))}{i \partial \nu}|_{\nu=0},
\end{align}
and a more simplified form can be obtained using RWA
\begin{align}\label{aveworksimplified}
	\ave{W}=(2A-1)\,\hbar\omega_0\,\sin ^2(G \Omega ),
\end{align}
indicating that the ground state absorbs work and the excited state releases work in amount of $ \hbar\omega_0$ in a sinusoidal way in harmony with the dynamics of the population or decoherency.

To find the work uncertainty, we use
\begin{align} \label{eq.second moment atom in ground state ,laser is on}
	\ave{W^2}&=\frac{\partial ^2 ch(\nu,t)}{ {(i)}^2 \partial \nu ^2},
\end{align}
leading to
\begin{align}
		 \ave{W^2}&=\frac{1}{4} \hbar ^2 \left(\alpha ^2+{\omega_0} ^2-2 {\omega_0} ^2 \cos (2 G {\Omega} )\right),
\end{align}
where we used RWA. Therefore,
\begin{align} \label{eq. variance of work atom in ground state ,laser is on}{({\Delta W})}=\frac{1}{2} \hbar  \sqrt{ \left( \left(\alpha ^2+{\omega_0} ^2-2 {\omega_0} ^2 \cos (2 G \Omega )\right)-4 {\omega_0} ^2 (2A-1)^2 \sin ^4 (G \Omega )\right)}.
\end{align}
The work distribution function now can be obtained from the inverse Fourier transform \cite{kheirandish2020many,talkner2007tasaki}
\begin{align}
\label{distribution function of work}
	P(W,t)&=	\int_{-\infty}^{+\infty} \frac{d\nu}{2 \pi}  e^{-i \nu \mathbf{W}}
	ch(\nu,\tau).
\end{align}
By inserting Eq. (\ref{def of charactristic RWA}) into Eq. (\ref{distribution function of work}), one finds
\begin{align}\label{wexchange}
	P(W,t)	&=\frac{1}{2} A  \left(1+\frac{{\omega_0} \cos (2 G \Omega )}{\alpha }\right)\boldsymbol{\delta} \left(W+\frac{1}{2} \hbar  (\alpha -{\omega_0} )\right)+\frac{1}{2} A  \left(1-\frac{{\omega_0} \cos (2 G \Omega )}{\alpha }\right) \boldsymbol{\boldsymbol{\delta}} \left(W-\frac{1}{2} (\alpha +{\omega_0} ) \hbar \right)
	\nonumber
	\\
&+\frac{1}{2} (1-A)  \left(1-\frac{{\omega_0} \cos (2 G \Omega )}{\alpha }\right) \boldsymbol{\delta} \left(W+\frac{1}{2} \hbar  (\alpha +{\omega_0} )\right)+\frac{1}{2}(1-A) \left(1+\frac{{\omega_0}  \cos (2 G \Omega )}{\alpha }\right) \boldsymbol{\delta} \left(W-\frac{1}{2} (\alpha -{\omega_0} ) \hbar \right).
\end{align}

From Eq. (\ref{wexchange}) it is seen that the work is exchanged only in the amounts of $(\mp\alpha+ \omega_0)\dfrac{\hbar}{2}$ with the probabilities
$
\frac{1}{2} A  \left(1\pm\frac{\omega _0 \cos (2 G \Omega )}{\alpha }\right)
$, and also in amounts of  $\dfrac{(- \alpha\pm \omega_0)\hbar}{2}$
with the probabilities of $\frac{1}{2} (1-A)  \left(1\pm\frac{\omega _0 \cos (2 G \Omega )}{\alpha }\right)$, respectively. The change of internal energy is defined by
\begin{align}
\text{$\Delta $u}(t)=\mbox{Tr}[H(t)\,\rho(t)]-\mbox{Tr}[H(0)\,\rho (0)],
\label{Eq.internal energy}
\end{align}
by inserting
\begin{align}
\label{matrix of H(t)}
H(t)=
 \left(
 \begin{array}{cc}
 	0 & -e^{i (\theta +\omega_0 t)} g\hbar \Omega    \\
 	-e^{-i (\theta + \omega_0 t)} g \hbar\Omega    &\hbar \omega _0  \\
 \end{array}
 \right),
 \end{align}
and Eq. (\ref{eq. rho_0 for atom in classiac laser}) into Eq. (\ref{Eq.internal energy}),
we will recover Eq. (\ref{aveworksimplified}) due to the conservation of energy in thermodynamics. If we take into account the fast oscillating terms in Eq.(\ref{Htilde2}), the averaged work is obtained as
\begin{align}
\ave{W}= & p(E_1)\, {|\bra{E_1}.\ket{E_1(t)}|}^2 (E_1(t)-E_1)+p(E_2)\, {|\bra{E_2}\ket{E_2(t)}|}^2 (E_2(t)-E_2)
\nonumber
\\
&+p(E_1)\, {|\bra{E_1}\ket{E_2(t)}|}^2 (E_2(t)-E_1)+p(E_2)\, {|\bra{E_2}\ket{E_1(t)}|}^2 (E_1(t)-E_2),
\end{align}
and for the characteristic function we obtain
\begin{align}
\label{def charactric of free energy helmholtz NP}
ch(\nu,t))= & p(E_1)\, {|\bra{E_1}\ket{E_1(t)}|}^2 e^{i \nu(E_1(t)-E_1)}+p(E_2)\, {|\bra{E_2}\ket{E_2(t)}|}^2  e^{i \nu(E_2(t)-E_2)}
\nonumber
\\
&+p(E_1)\, {|\bra{E_1}\ket{E_2(t)}|}^2  e^{i \nu(E_2(t)-E_1)}+p(E_2)\, {|\bra{E_2}\ket{E_1(t)}|}^2  e^{i \nu(E_1(t)-E_2)},
\end{align}
where $E_j(t)$ are the eigenvalues of
\begin{align}
\label{matrix of H(t)}
H_{tot}(t)=
\left(
\begin{array}{cc}
0 & -e^{i \theta} g\hbar \,\Omega   \,\cos{\omega_0 t }  \\
-e^{-i \theta} g \hbar
\Omega  \,\cos{\omega_0 t }  & \omega_0  \hbar  \\
\end{array}
\right),
\end{align}
with the corresponding eigenstates
$\ket{E_j(t)}$,
respectively. Note that $E_j$'s are the eigenvalues of the initial Hamiltonian with the eigenstates $\ket{E_j}$ and the corresponding probabilities $P(E_j),\,\,(j=1,2)$.
\subsubsection{The Helmoholtz free energy}
The variation of the free energy, $\Delta F$=  $F(T,t_f)-F(T,t_i,)$, is the difference between the free energy at final and initial equilibrium states
with Hamiltonians $H(t_f)$ and $H(t_i)$, at the same temperature $T$, respectively. It encodes the response of a system to the variation of a time-dependent  Hamiltonian. By making use of the Jarzynski theorem we find
\begin{align}
   	\label{def of helmholta free energy}
   \Delta F(T,t)  =\frac{-1}{\beta} \text{Log}\left[ ch(i \beta,t)\right],
\end{align}
where $\beta=1/K_B T$ and $K_B$ is the Boltzman's constant. By inserting Eq. (\ref{def of charactristic RWA}) into Eq. (\ref{def of helmholta free energy}), and using RWA, we will find
\begin{figure}[!tp]
	\centering
	\begin{subfigure}[b]{0.48\textwidth}
		\centering
	\includegraphics[width=\textwidth]{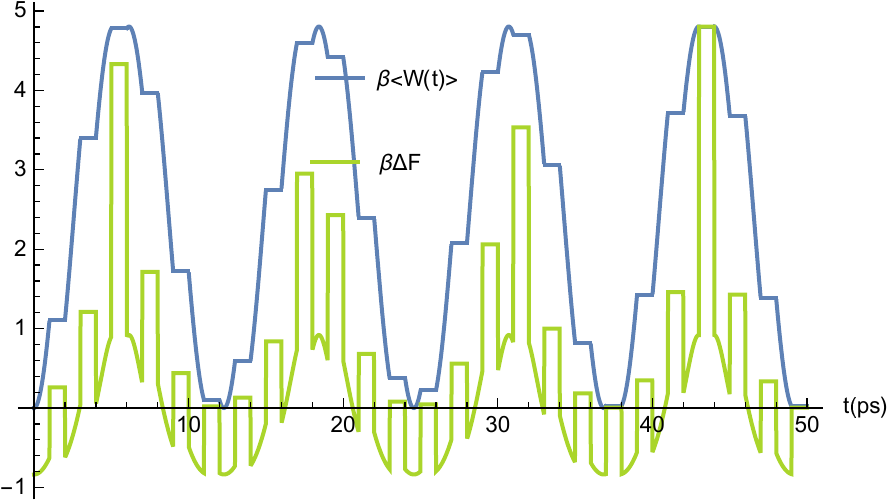}
		\caption{If we assume RWA is valid}
		\label{FigsFW}
	\end{subfigure}
	\hfill
	\begin{subfigure}[b]{0.48\textwidth}
		\centering
		\includegraphics[width=\textwidth]{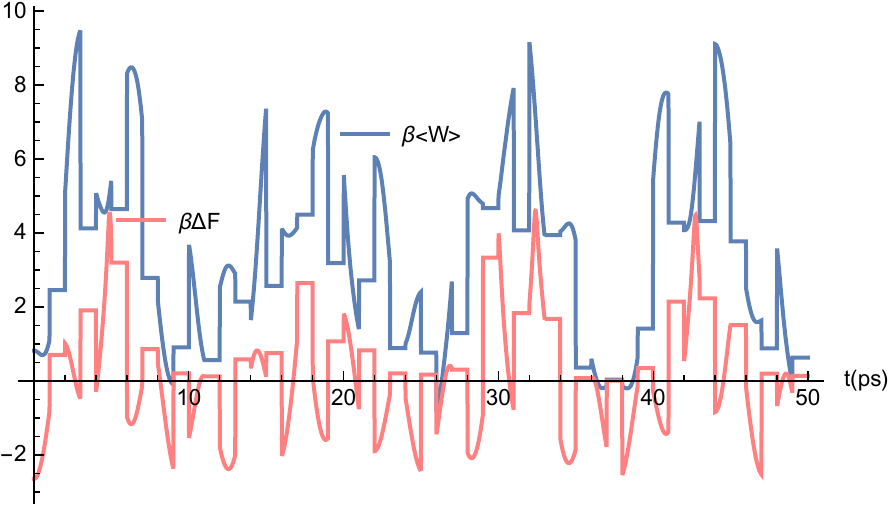}
		\caption{Without assuming RWA and keeping all terms in Eq. (\ref{Htilde2})}
\label{FigsFWNd}
	\end{subfigure}
	\hfill
	\caption{The average work and variation of the Helmholtz free energy both scaled by $\beta$, are depicted in time (picosecond) when the atom is initially prepared in the ground state. The parameters are chosen as $\Omega=.5$ \mbox{teraHz}, $\omega=1$ \mbox{teraHz}, $\xi=0$, and $\theta=0$, $T=10K^\circ$.}
	\label{popi and decoh}
\end{figure}
\begin{align}
	\Delta F =& -\text{Log}\left[\frac{\omega  \cos (2 G \Omega ) \sinh (\frac{\alpha  \beta  \hbar }{2}) \left((2 A-1) \cosh (\frac{\beta  \omega  \hbar }{2})-\sinh (\frac{\beta  \omega  \hbar }{2})\right)}{\alpha }\right.
\nonumber	
\\
&\left.+\cosh (\frac{\alpha  \beta  \hbar }{2}) \left((1-2 A) \sinh (\frac{\beta  \omega  \hbar }{2})+\cosh (\frac{\beta  \omega  \hbar }{2})\right)\right].	
		\label{eq.Hel of James }
\end{align}
A more accurate result can be obtained by inserting Eq. (\ref{def charactric of free energy helmholtz NP})
into Eq. (\ref{def of helmholta free energy}) which is depicted in Fig. (\ref{FigsFWNd}).

Helmholtz free energy is intensively dependent on temperature. If temperature increases, the lower curves tend to overlap the upper curves which is consistent with the classical thermodynamic and also from Figs. (\ref{popi and decoh}) we see that the Jarzynski theorem $W_{irr}=\ave{W} - \Delta F\geq 0$ is verified,  \cite{jarzynski2007comparison,jarzynski1997nonequilibrium,plastina2014irreversible}.
\section{Two-level atom trapped in a harmonic potential interacting with a classical light field}\label{generalisation to vibrating atom}
\label{vibronic-atom trapped}
In this section, we will evaluate the level occupation probabilities and work distribution function for an oscillating two-level atom interacting with a classical laser field \cite{lipfert2018time}. The Hamiltonian describing this system can be written as
\begin{align} \label{hamiltonian of nonlinear Jaynes-Cummings dynamics of a trapped ion   }
\hat{H}=\hbar \nu a^\dagger a +\hbar \omega_{21}\sigma^+  \sigma^-+
g(t)\left(\hbar   {|\kappa| } \sigma^+ e^{-i\omega_l t} \cos[\eta(a+a^\dagger)]+h.c.\right),
\end{align}
where $g(t)$ is a switching function that controls the duration of interaction ($\tau$) between the system and laser field. Therefore, $g(t)=0$ for $t>\tau$. From now on we drop the argument of g(t) for notational simplicity.
The operators $ a^\dagger$ $ (a) $ are the creation(annihilation) operators of the vibrational degrees of freedom of the atom with frequency $\nu $. The frequency $\omega_{21}$ denotes the atomic transition frequency, the operators $\sigma^+ =|2\ra\la 1|$ and $ \sigma^- =|1\ra\la 2|$ are atomic transition operators, $\eta$ denotes the Lambe-Dicke parameter and $ {|\kappa| }$ denotes the strength of the coupling between the electric atomic dipole and the electric component of the laser field \cite{lipfert2018time}.
%%%%%%%%%%%%%%%%%%%%%%%%%%%%%%%%%%%%%%%%%%%%%%%%%%%%%%%%%%%%%%%%%%%%%%%%%%%%%%%%%%%%%%%%%%%%%%%%%%%%%%%%%%%%%%%%%%%%%%%%%%%%%%%%%%%%%%%%%%%%%%%%%%%%%%%%%%%%%%%%%%%%%%%%%%%%%%%%%%%%%%
\subsection{Level-occupation probabilities}
%%%%%%%%%%%%%%%%%%%%%%%%%%%%%%%%%%%%%%%%%%%%%%%%%%%%%%%%%%%%%%%%%%%%%%%%%%%%%%%%%%%%%%%%%%%%%%%%%%%%%%%%%%%%%%%%%%%%%%%%%%%%%%%%%%%%%%%%%%%%%%%%%%%%%%%%%%%%%%%%%%%%%%%%%%%%%%%%%%%%%%
Let us assume that the combined system is initially prepared in the product state
\begin{align}
 \label{Eq.ro0 in vibronic model }
     \rho(0)= \sum_{n=0}^{\infty}p(n) \ket{n } \bra{n } \otimes\left(\mathbb{A}\ket{2}\bra{2}+\mathbb{B}\ket{1}\bra{1}\right),
\end{align}
where $n$ belong to the complete set of eigenstates in the Hilbert space of the vibrational motion, and $\ket{1}, \ket{2}$ are the ground and excited states of the atomic Hilbert space. We denote the occupation probability of the $n$th eigenstate of the vibrational motion by $p(n)$ and let $\mathbb{A}$ and $\mathbb{B}$ be the occupation probabilities corresponding to the excited and ground state of the atomic Hilbert space, respectively. The population probability is
\begin{align}
\label{Eq.popj for vibronic model}
popj(\tau)=\bra{j}U(t,0)\,\rho_0\,U^\dagger(t,0)\ket{j},\,\,\,(j=2, 1),
\end{align}
By inserting Eqs. (\ref{U in vibronic model}) into Eq. (\ref{Eq.popj for vibronic model}) the population dynamics of the excited state is obtained as
\begin{align}
pop2(\tau)=\mathbb{A} \sum_{n=0}^{\infty}{|a_{n}|} ^2 p(n)+\mathbb{B} \sum_{n=0}^{\infty}{|b_{n}|} ^2 p(n+k),
\label{pop2}
\end{align}
where the laser is slightly detuned from the $k$th sideband (App. \ref{appendixC}). Similarly, for the ground state we find
\begin{align}
pop1(\tau)=\mathbb{A} \sum_{n=0}^{\infty}{|b_{n}|} ^2 p(n)+\mathbb{B} \sum_{\boldsymbol{n}=0}^{\infty}{|a_{n}|} ^2 p(n+k)+\mathbb{B} \sum_{n=0}^{k-1}p(n),
\label{{pop1}}
\end{align}
where
\begin{align}
& a_{n}(\tau,o)= e^{-i  \frac{\Delta \omega \tau} {2}} [\cos(\Gamma_{n}\tau)+ \frac{i\Delta \omega}{2 \Gamma_{n}}  \sin(\Gamma_{n}\tau)],\,\,\,n=0,1,2,\cdots,\\
& b_{n}=  e^{-i  \frac{\Delta \omega \tau} {2}} \frac{|\kappa|(g \, \omega_{n})}{i \Gamma_n}  \sin(\Gamma_n\tau),\\
& \Gamma_n=\sqrt{{\frac{\Delta \omega}{2} }^2 +{g}^2 \omega^2_n   {|\kappa| }^2},\\
& \omega_{n}=\cos(\frac{\pi k}{2})  \, \eta^k \, e^{  \frac{-\eta^2}{2}}   \sqrt{ \frac{n!}{(n+k)!}} \, L_{n}^{(k)} ( \eta^2),\\
& \Delta \omega =\omega_L -\omega_{21}+k \nu,\,\,\,(\Delta \omega \ll \nu),
\end{align}
and $L_n^{(k)} (\eta^2)$ is the modified Laugure polynomial \cite{kheirandish2020many,talkner2007tasaki,lipfert2018time} (App. (\ref{appendixB})).
\begin{figure}[!tp]
	\centering
	\begin{subfigure}[b]{0.48\textwidth}
		\centering
		\includegraphics[width=\textwidth]{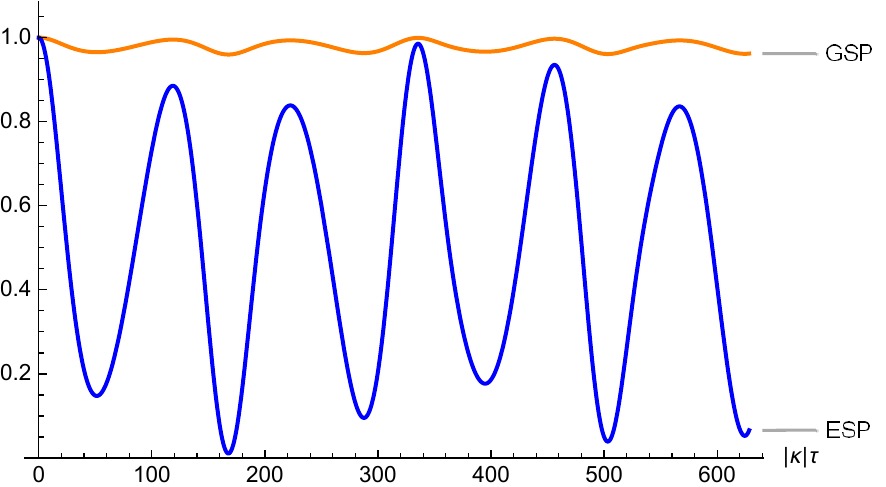}
		\caption{$T=30k^\circ$}
		\label{FigpopGET1}
	\end{subfigure}
	\hfill
	\begin{subfigure}[b]{0.48\textwidth}
		\centering
		\includegraphics[width=\textwidth]{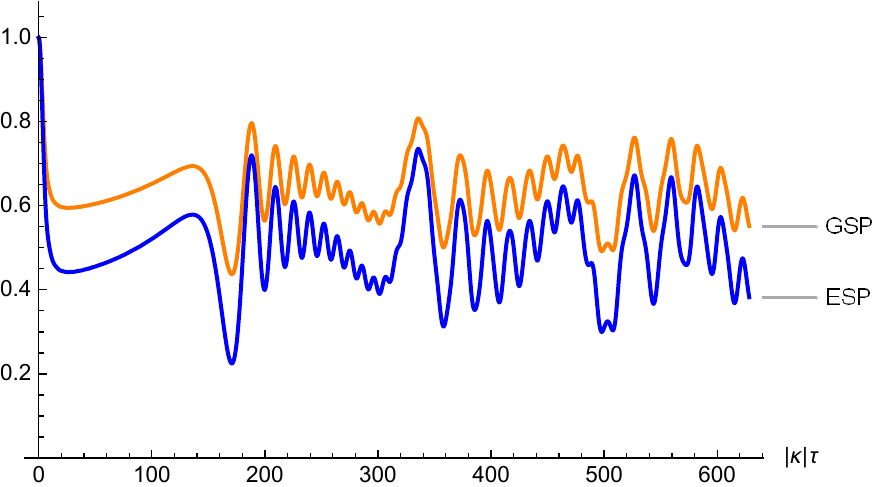}
		\caption{$T=300k^\circ$}
		\label{FigpopGET2}
	\end{subfigure}
	\hfill
	
	\caption{The ground state occupation probability (\mbox{GSP}) when it is initially occupied and the excited state occupation probability (\mbox{ESP}) when it is initially occupied are depicted in dimensionless parameter $|\kappa|\,\tau$ for a trapped two-level atom in an external classical light field for temperatures $T=30k^\circ$ (\mbox{Left}) and $T=300k^\circ$ \mbox{Right}. The parameters are chosen as $\Delta\omega/|\kappa|=0.005,\, k=2,\, \eta=.2,\, \theta=0$, and $g(t)=\Theta(t)\,\Theta(\tau-t)$. $\Theta(t)$ is the Heaviside step function.}
		\label{FigpopEG}
\end{figure}	
%%%%%%%%%%%%%%%%%%%%%%%%%%%%%%%%%%%%%%%%%%%%%%%%%%%%%%%%%%%%%%%%%%%%%%%%%%%%
In Fig. (\ref{FigpopEG}), the curve GSP denotes the dynamics of the population of the ground state where it has been initially occupied and  the curve ESP denotes the population dynamics of the excited state, when it has been initially occupied.

Fig. (\ref{FigpopGET1}) demonstrates the population dynamics of the ground and excited states at the temperature $T=30k^\circ$. The orange curve illustrates that the ground state has been occupied initially. It demonstrates that the reaction of the ground state is slow and very inert compared to the excited state, which is fast and agile. It is expectable because sufficient energy is needed to jump up the electron. Comparing Fig. (\ref{FigpopGET1}) with Fig. (\ref{FigpopGET2}), we see that both curves almost behave in the same way for higher temperatures. Also, the increase of temperature cause an increase of the decoherency between electronic states since increasing the temperature stimulates more harmonics(Phonons) and higher level states will be populated with nonzero probabilities regardless of the initial state be occupied. Furthermore, the curves show the occurrence of sequential collapses and revivals with growing time.
%%%%%%%%%%%%%%%%%%%%%%%%%%%%%%%%%%%%%%%%%%%%%%%%%%%%%%%%%%%%%%%%%%%%%%%%%%%%%%%%%%%%%%%%%%%%%%%%%%%%%%%%%%%%%%%%%%%%%%%%%%%%%%%%%%%%%%%%%%%%%%%%%%%%%%%%%%%%%%%%%%%%%%%%%%%%%%%%%%%%%%
\subsection{The characteristic function}
%%%%%%%%%%%%%%%%%%%%%%%%%%%%%%%%%%%%%%%%%%%%%%%%%%%%%%%%%%%%%%%%%%%%%%%%%%%%%%%%%%%%%%%%%%%%%%%%%%%%%%%%%%%%%%%%%%%%%%%%%%%%%%%%%%%%%%%%%%%%%%%%%%%%%%%%%%%%%%%%%%%%%%%%%%%%%%%%%%%%%%
The characteristic function $ch(\lambda,t)$ \cite{kheirandish2020many,talkner2007tasaki} of the trapped two-level atom interacting with a classical driving laser at the times $t> \tau$ can be obtained from Eq. (\ref{definition of charastric function for a two level atom}), (App.~\ref{appendixC})
\begin{align}
	\label{Eq.Gvib }
ch(\lambda,\tau)=&\sum_{{n}=0}^{\infty}  \left(
	{|a_{n}| }^2+{|b_{n}| }^2 e^{-i \lambda \left(\hbar \omega_{21   }- k \hbar \nu  \right)}
	\right)\,\bra{2,{n}} \rho(0) \ket{2,n}
	+\sum_{{n}=0}^{\infty} a_{n} b_{n}^* \left(
	e^{i \lambda  \left(\hbar \omega_{21} - k \hbar \nu \right) }
	-1\right)
	\,\bra{ 2,n  }\rho(0) \ket{1,n+k }\nonumber \\
	+&\sum_{n=0}^{\infty}
	a_{n}^*  b_{n} \left(1-
	e^{-i \lambda  \left(\hbar \omega_{21} - k \hbar \nu \right) }
\right)
	\,\bra{ 1,n+k  } \rho(0)  \ket{2,{n} }
	+\sum_{n=0}^{\infty}
	\left(
	{|a_n|}^2+{|b_{n}|}^2 e ^{i \lambda \left(
		\hbar \omega_{21}- k \hbar \nu	
		\right)}
	\right) \bra{ 1,n+k}\rho(0)\ket{1,{n}+k }
	\nonumber
	\\+& \sum_{n=0}^{k-1} \bra{1,n}\rho(0)\ket{1,n},
\end{align}
where $n$ refers to the $n$th eigenstate of the vibrational motion and numbers $1$, $2$ refer to the ground and excited states, respectively. If we insert the initial density matrix of the combined system from Eq. (\ref{Eq.ro0 in vibronic model }) into Eq. (\ref{Eq.Gvib }), we will find (App. \ref{appendixB})
\begin{align}\label{charac}
ch(\lambda ,\tau )=&  \sum_{n=0}^{\infty}\Big(\mathbb{A}\,p(n)  e^{-i \lambda (\hbar \omega_{21}-k\hbar\nu)}+
\mathbb{B} p(n+k) e^{i \lambda (\hbar \omega_{21   }-k\hbar \nu )} \Big){|b_{n}|}^2,
\nonumber
\\
+&
\sum_{n=0}^{\infty} \Big(\mathbb{A} p(n) +
\mathbb{B} p(n+k)\, \Big){|a_{n}|}^2 +
\mathbb{B} \sum_{n=0}^{k-1}\,p(n).
\end{align}
Having the characteristic function, we find the average work $\ave{W}$ done on the vibronic two-level atom as
\begin{figure}[tp]
	\centering
	\begin{subfigure}[b]{0.48\textwidth}
		\centering
		\includegraphics[width=\textwidth]{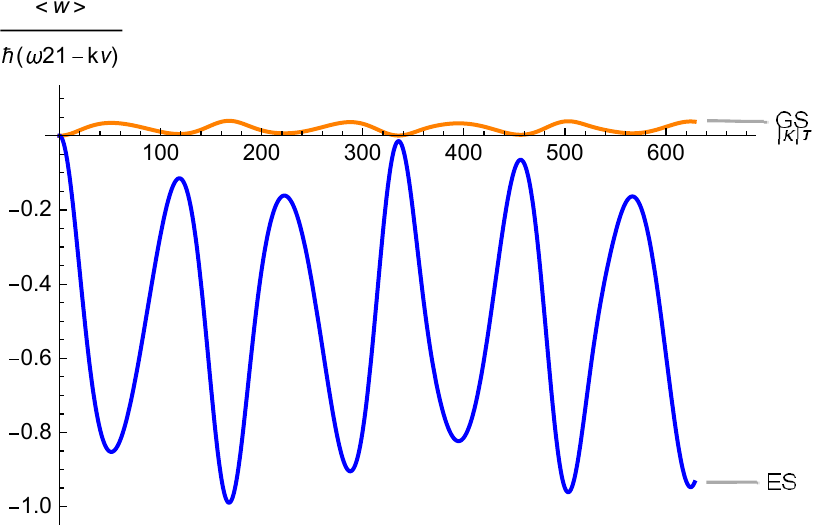}
		\caption{$T=30k^\circ$}
		\label{wT1}
	\end{subfigure}
	\hfil
	\begin{subfigure}[b]{0.48\textwidth}
		\centering
	\includegraphics[width=\textwidth]{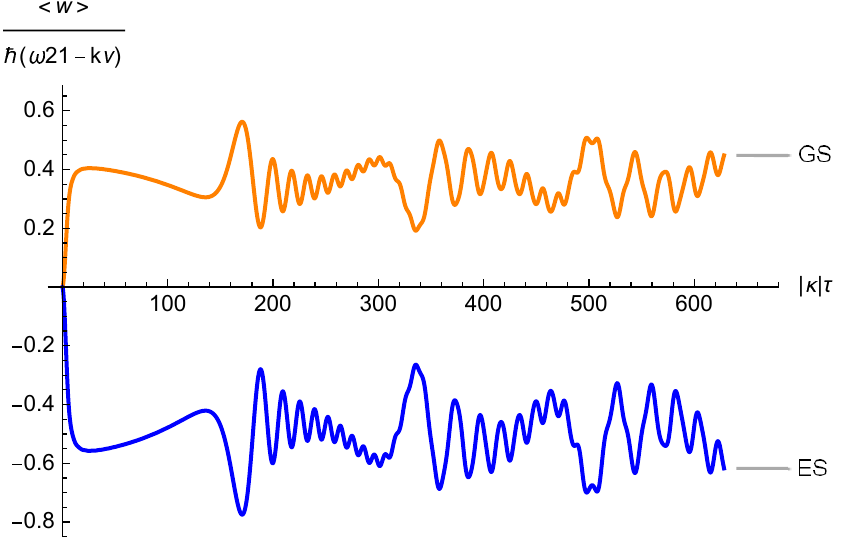}
		\caption{$T=300k^\circ$}
		\label{Fig.wT2}
	\end{subfigure}
	\hfill
	\caption{The scaled average work from Eq.~(\ref{work of vib}) versus the dimensionless parameter $|\kappa|\,\tau$ (scaled duration of interaction). The curve (\mbox{ES}) refers to initially occupied excited state and the curve (\mbox{GS}) to the initially occupied ground state. The parameters are chosen as $\Delta\,\omega/|\kappa|=.005,\, k=2$, $\eta=.2$, $\theta=0$, and $g(t)=\Theta(t)\,\Theta(\tau-t)$.}
		\label{Fig.work of vib }
\end{figure}
\begin{align}
\ave{W} =\mathbb{A}(k \hbar \nu -\hbar \omega_{21})\,\sum_{n=0}^{\infty} p(n)\,{|b_n|}^2 +\mathbb{B}( \hbar \omega_{21}-k \hbar \nu) \sum_{n=0}^{\infty}p(n+k)\,{|b_n|}^2.
\label{work of vib}
\end{align}
From Eq. (\ref{work of vib}) we observe that the oscillator loses the amount of energy $\hbar \omega_{21}-k \hbar \nu $ with the probability $p(n)\,{|b_n|}^2$ if it is initially prepared in the state $\ket{2,n}$. Also, the oscillator attains the same amount of energy with the probability $ p(n+k)\,{|b_n|}^2 $ if it is initially prepared in the state $\ket{1,n}$, see Fig. (\ref{Fig.work of vib }). In Figs. (\ref{wT1}, \ref{Fig.wT2}), the scaled average work $\ave{W}$ is depicted in dimensionless parameter $|\kappa|\,\tau $ at the temperatures $T=30k^\circ$, $T=300k^\circ$ for both atomic initial states $\ket{2}$ (ES) and $\ket{1}$ (GS). Furthermore, if we pursue to plot at higher temperatures, the steady value approaches to $0.5$ the more and the domain of fluctuations becomes lesser, which is compatible with the theorem of the equipartition energy in classical thermodynamics and the indistinguishably between states is negligible.
%%%%%%%%%%%%%%%%%%%%%%%%%%%%%%%%%%%%%%%%%%%%%%%%%%%%%%%%%%%%%%%%%%%%%%%%%%%%%%%%%%%%%%%%%%%%%%%%%%%%%%%%%%%%%%%%%%%%%%%%%%%%%%%%%%%%%%%%%%%%%%%%%%%%%%%%%%%%%%%%%%%%%%%%%%%%%%%%%%
\subsection {The work distribution function}
%%%%%%%%%%%%%%%%%%%%%%%%%%%%%%%%%%%%%%%%%%%%%%%%%%%%%%%%%%%%%%%%%%%%%%%%%%%%%%%%%%%%%%%%%%%%%%%%%%%%%%%%%%%%%%%%%%%%%%%%%%%%%%%%%%%%%%%%%%%%%%%%%%%%%%%%%%%%%%%%%%%%%%%%%%%%%%%%%
By inserting Eq. (\ref{charac}) into Eq. ( \ref{distribution function of work}), we find the work distribution function as
\begin{align} \label{eq.distribution of work for vibronic atom}
	p(W)=&
\boldsymbol{\delta}\left(W+ \hbar \omega_{21}-k \hbar \nu    \right)\,	\mathbb{A}\sum_{n=0}^{\infty}\,p(n)  {|b_{n}|}^2+
	\boldsymbol{\delta}(W)\,\mathbb{A}\sum_{n=0}^{\infty}  p(n){|a_{n}|}^2
	 \nonumber \vspace{.6cm} \\
	+ & \boldsymbol{\delta}\left(W -\hbar \omega_{21}+k \hbar \nu\right)\,\mathbb{B}\sum_{n=0}^{\infty} p(n+k) {|b_{n}|}^2+\boldsymbol{\delta}(W)\,\mathbb{B}\sum_{n=0}^{\infty}  p(n+k) {|a_{n}|}^2+\boldsymbol{\delta}(W)\,\mathbb{B}\sum_{n=0}^{k-1}\,p(n).
\end{align}
From Eq. (\ref{eq.distribution of work for vibronic atom}) we deduce that the vibronic atom loses(gains) the amount of energy $\hbar\,\omega_{21}-k \hbar\,\nu$ with the
probability $\sum\limits_{n=0}^{\infty} p(n){|b_n|}^2$ ($\sum\limits_{n=0}^{\infty} p(n+k){|b_n|}^2$) if it is initially prepared in the state $\ket{2}$ ($\ket{1}$), respectively. Also, the probability that no work be done on the atom, i.e the atom be transparent to the laser field is $\sum\limits_{n=0}^{\infty} p(n){|a_n|}^2$ ($\sum\limits_{n=0}^{\infty} p(n+k){|a_n|}^2+\sum\limits_{n=0}^{k-1}\,p(n)$) for the initial state $\ket{2}$ ($\ket{1}$), respectively.
%%%%%%%%%%%%%%%%%%%%%%%%%%%%%%%%%%%%%%%%%%%%%%%%%%%%%%%%%%%%%%%%%%%%%%%%%%%%%%%%%%%%%%%%%%%%%%%%%%%%%%%%%%%%%%%%%%%%%%%%%%%%%%%%%%%%%%%%%%%%%%%%%%%%%%%%%%%%%%%%%%%%%%%%%%
\section{conclusion}\label{conclusion}
%%%%%%%%%%%%%%%%%%%%%%%%%%%%%%%%%%%%%%%%%%%%%%%%%%%%%%%%%%%%%%%%%%%%%%%%%%%%%%%%%%%%%%%%%%%%%%%%%%%%%%%%%%%%%%%%%%%%%%%%%%%%%%%%%%%%%%%%%%%%%%%%%%%%%%%%%%%%%%%%%%%%%%%%%%
In Sec. \ref{Atom-light-classical}, the level-occupation probabilities of the ground and excited states of a two-level atom interacting with a classical light field controlled by a certain switching function was obtained analytically by assuming RWA and numerically without assuming RWA, Fig. 2. The decoherency function was obtained for both cases and it was shown in Fig. 3 that in the exact treatment the fluctuations of decoherency was lesser due to taking into account the high frequency terms.

The characteristic function and Helmholtz free energy were obtained for a two-level atom under a controlled light field. The averaged work received by the atom initially prepared in the ground state, was calculated and it was shown that the Jarzynski theorem is fulfilled in both cases (with and without RWA), Fig. 4.

In Sec. \ref{generalisation to vibrating atom}, we generalised the problem to the case of a trapped two-level atom in a harmonic potential and interacting with a classical light field controlled by a certain switching function at low temperature regime. The probabilities of occupation for the ground and excited states of the two-level atom were obtained and it was illustrated that the ground state population varies much lesser than the excited state so it was much more stable than the excited state, nevertheless, they overlap at sufficiently high temperatures manifesting the equal probability of occupations(ergodicity). The characteristic function and work distribution function were obtained and discussed for a trapped two-level atom under the influence of a classical light field. Finally, the validation of the Jarzynski theorem was illustrated.
%\textbf{Acknowledgment}..
	%========================================================================
	%========================================================================
	%========================================================================
\newpage
	
\appendix	
	
\numberwithin{equation}{section}	
%%%%%%%%%%%%%%%%%%%%%%%%%%%%%%%%%%%%%%%%%%%%%%%%%%%%%%%%%%%%%%%%%%%%%%%%%%%%%%%%%%%%%%%%%%%%%%%%%
%========================================================================
%========================================================================
%==================================\newpage
\appendix	
\numberwithin{equation}{section}	
%\newpage
%%%%%%%%%%%%%%%%%%%%%%%%%%%%%%%%%%%%%%%%%%%%%%%%%%%%%%%%%%%%%%%%%%%%%%%%%%%%%%%%%%%%%%%%%%%%%%%%%%
\section{} \label{appendixA}
%\newpage
The time-evolution operator for the Jaynes-Cummings model in the off-resonance case fulfills the equation
\begin{align}
\label{Eq.Scrodinger equation}
\tilde{H} \tilde{U}=i\hbar\frac{d\tilde{U}}{dt},
\end{align}
by inserting the equation
\begin{align}
 \tilde{U}(t)=e^{-i \Delta \, t \sigma^+ \sigma^- }\; \tilde{\hat{V}}(t),
 \end{align}
and Eq. (\ref{Htilde2}) into Eq. (\ref{Eq.Scrodinger equation}), we find
\begin{align}
\hbar \Delta \sigma^+ \sigma^- \tilde{U}(t)+i \hbar e^{-i \Delta t \sigma^+ \sigma^-} \;
 \frac{d\tilde{\hat{V}}(t)}{dt}
=\hbar\Delta  \sigma^+ \sigma^-\tilde{U}(t)-g\hbar \Omega \left(e^{-i \theta} \sigma^++h.c\right) \tilde{U}(t),
\end{align}
therefore,
\begin{align}
\label{dvtilde}
\frac{d\hat{\tilde{V}}(t)}{dt}&=ig \Omega e^{i \Delta t \sigma^+ \sigma^-}\left(e^{-i \theta} \sigma^++h.c.\right)
  e^{-i\Delta \,t \sigma^+ \sigma^-}
  \hat{\tilde{V}}(t), \nonumber \\
  &=ig \Omega   \left(e^{-i( \theta-\Delta t)}  \sigma^+ + h.c.\right)
    \hat{\tilde{V}}(t).
  \end{align}
By assuming the following form for $\hat{\tilde{V}}(t)$
\begin{align}
\hat{\tilde{V}}(t)=e^{i A(t) \sigma^-} e^{i B(t) \sigma^+}
e^{i C(t) \sigma^z} e^{i f(t) },
\end{align}
and inserting it into Eq. (\ref{dvtilde}), we will find the following equations to determine the unknown coefficients
\begin{align}
 	 \dot{B}A+i\dot{C}-2iB \dot{C}&=0,
 	\\
-i \dot{A}-i\dot{B} A^2+2A\dot{C}-2 \dot{C} BA^2&=ig\Omega e^{-i\Delta t}
e^{i\theta },
\\
 i \dot{B}+2B \dot{C}&=  ig\Omega e^{i\Delta t}
 e^{-i\theta },
\\
\dot{f}(t)&=0.
  \end{align}
%%%%%%%%%%%%%%%%%%%%%%%%%%%%%%%%%%%%%%%%%%%%%%%%%%%%%%%%%%%%%%%%%%%%%%%%%%%%%%%%%%%%%%%%%%%%%%%%%%%%%%%%%%%%%%%%%%%%%%%
\newpage
%%%%%%%%%%%%%%%%%%%%%%%%%%%%%%%%%%%%%%%%%%%%%%%%%%%%%%%%%%%%%%%%%%%%%%%%%%%%%%%%%%%%%%%%%%%%%%%%%%%%%%%%%%%%%%%%%%%%%%%
\section{} \label{appendixB}

%%%%%%%%%%%%%%%%%%%%%%%%%%%%%%%%%%%%%%%%%%%%%%%%%%%%%%%%%%%%%%%%%%%%%%%%%%%%%%%%%%%%%%%%%%%%%%%%%%%%%%%%%%%%%%%%%%%%%%%
We have
\begin{align}
\label{   }
 \cos[\eta(a+a^\dagger)]=\dfrac{1}{2}e^{\frac{–\eta^2}{2}} \sum_{l,m=0}^\infty
 \frac{{(i\eta)}^{l+m}}{l!m!} {a^\dagger}^la^m + h.c.,
\end{align}
The Hamiltonian in the interaction picture(indicated by tilde), reads as
\begin{align}
\label{   }
\tilde{H}_{int}
=\frac{1}{2}\hbar|\kappa| \sigma^+e^{\frac{–\eta^2}{2}} \sum_{l,m=0}^\infty
\frac{{a^\dagger}^la^m}{l!m!} e^{-i\left[{\omega_L -\omega_{21   }+(m-l)\nu}\right]t}
\big[{(i\eta)}^{l+m}+{(-i\eta)}^{l+m}\big]+h.c.
\end{align}
For the purpose of the present paper, we are interested in a situation when the laser is slightly detuned from the $k$th sideband
\begin{align}
	\omega_L =\omega_{21}-k\nu+\Delta\omega,
		\end{align}
with $\Delta\omega\ll\nu$. For more details see \cite{lipfert2018time}.
%%%%%%%%%%%%%%%%%%%%%%%%%%%%%%%%%%%%%%%%%%%%%%%%%%%%%%%%%%%%%%%%%%%%%%%%%%%%%%%%%%%%%%%%%%%%%%%%%%%%%%%%%%%%%%%%%%%%%%%%%%%%%%%%%%%%%%
\newpage
%%%%%%%%%%%%%%%%%%%%%%%%%%%%%%%%%%%%%%%%%%%%%%%%%%%%%%%%%%%%%%%%%%%%%%%%%%%%%%%%%%%%%%%%%%%%%%%%%%%%%%%%%%%%%%%%%%%%%%%%%%%%%%%%%%%%%%
\section{} \label{appendixC}
%%%%%%%%%%%%%%%%%%%%%%%%%%%%%%%%%%%%%%%%%%%%%%%%%%%%%%%%%%%%%%%%%%%%%%%%%%%%%%%%%%%%%%%%%%%%%%%%%%%%%%%%%%%%%%%%%%%%%%%%%%%%%%%%%%%%%%
The evolution-operator of a vibronic atom interacting with a classical laser field with an arbitrary switching function G, can be represented in the interaction picture as \cite{lipfert2018time}
\begin{align}
\tilde{U}(t,0)=\sum_{n=0}^{\infty} \psi_{n}^\dagger  U_{n}(t,0) \psi_{n} + \sum_{q=0	}^{k-1}|1,q \rangle \langle1,q|,
\label{tilde U}
\end{align}
where $U_{n}(t,0)\in C^{2 \times 2}, (n=0,1,2,\cdots)$ is given by
\begin{align}
	\label{definitions of an and bn}
	{U}_{n}(t,0)=
	\begin{bmatrix}
	a_{n}&  b_{n}(t,0) \\   -b_{n}^*(t,0)& a_{n}^*(t,0)
	\end{bmatrix},
\end{align}
and ${|a_{n}(t,0)|}^2+  {|b_{n}(t,0)|}^2=1$. The coefficients $a_{n}$ and $b_{n}$ are given by \cite{lipfert2018time}
\begin{align}
	a_{n}(t,0)=e^{-i \Delta \omega (\frac{t}{2})}
	[ \cos(\Gamma_{n}t)+\dfrac{i \Delta \omega}{2 \Gamma_n} \sin(\Gamma_{n}t)],\\
%	\qquad
	b_{n}(t,0)=e^{-i \Delta \omega (\frac{t}{2})}
	\dfrac{|\kappa| g\, \omega_{n}}{i \Gamma_{n}} \sin(\Gamma_{n}t),
\end{align}
where
\begin{align}
\Gamma_{n}=\sqrt{{(\dfrac{\Delta \omega}{2})}^2+g^2 \omega_{n}^2  {|\kappa_{n}|}^2},
\end{align}
\begin{align}
\omega_{n}=\cos(\dfrac{ \pi k}{2}) \, \eta^k \, e^{- \frac{\eta^2}{2}}
\sqrt{\dfrac{n!}{(n+k)!}}\, L_{n}^{(k)}(\eta^2),
\end{align}
and the spinors
\begin{align}
	\psi_{n}=
	\begin{bmatrix}
	\bra{2,{n}} e^{-i \frac{\theta}{2}}
	\\
	\bra{1,n+k} e^{i \frac{\theta}{2}}
	\end{bmatrix},
\end{align}
fulfill $\psi_{n}\psi_{n^\prime}^\dagger =I \delta_{n,n^\prime}$ where
$	I=
\begin{bmatrix}
1&  0 \\
0& 1
\end{bmatrix}$,
for more details see \cite{lipfert2018time}.

The evolution-operator can be obtained from
\begin{align}
	\label{U in vibronic model}
	 U(t,0)=U_0(t,0)\tilde{U}(t,0),
\end{align}
where $ U_0(t,0)$ is the free evolution-operator defined by
\begin{align}
	 U_0(t,o)&=e^{-\dfrac{i t}{\hbar} H_0(0)},
	 \nonumber
	 \\
	 &=
	e^{-i \left(n \nu +\omega_{21}\right)t }\ket{2,n}\bra{2,n}  +
e^{-i \left( n+k\right) \nu t}\ket{1,n}\bra{1,n}.
	\label{U_0 for vibronic}
\end{align}
The characteristic function is defined by
\begin{align}
 ch(\lambda,t)= Tr\left(e^{- i\lambda H_0} U^\dagger(t,0) e^{i \lambda H} U(t,0)\rho(0) \right),
\end{align}
therefore, at times where the switching function is turned off ($t>\tau$), we have
\begin{align}
ch(\lambda,t)=Tr\left(e^{-i \lambda H_0} \tilde{U}^\dagger(\tau,o)
	e^{i \lambda H_0}  \tilde{U}(\tau,o)\rho(0) \right).
\end{align}
%%%%%%%%%%%%%%%%%%%%%%%%%%%%%%%%%%%%%%%%%%%%%%%%%%%%%%%%%%%%%%%%%%%%%%%%%%%%%%%%%%%%%%%%%%%%%%%%%%%%%%%%%%%%%%%%%%%%%%%%%%%%%%%%%%%%%%%%%%%%%%%%%%%%%%%%%%%%%%%%%%%%%%%%%%%%%
\newpage
%%%%%%%%%%%%%%%%%%%%%%%%%%%%%%%%%%%%%%%%%%%%%%%%%%%%%%%%%%%%%%%%%%%%%%%%%%%%%%%%%%%%%%%%%%%%%%%%%%%%%%%%%%%%%%%%%%%%%%%%%%%%%%%%%%%%%%%%%%%%%%%%%%%%%%%%%%%%%%%%%%%%%%%%%%%%%
\bibliography{QThermoTrap-EPJP}

\begin{thebibliography}{10}

\bibitem{tasaki2000jarzynski}
H.~Tasaki.
\newblock Jarzynski relations for quantum systems and some applications.
\newblock {\em arXiv preprint cond-mat/0009244}, 2000.

\bibitem{Roeck2004quantum}
W.~De~Roeck and C.~Maes.
\newblock Quantum version of free-energy--irreversible-work relations.
\newblock {\em Physical Review E}, 69(2):026115, 2004.

\bibitem{talkner2007tasaki}
P.~Talkner and P.~H{\"a}nggi.
\newblock The tasaki--crooks quantum fluctuation theorem.
\newblock {\em Journal of Physics A: Mathematical and Theoretical},
  40(26):F569, 2007.

\bibitem{deffner2008nonequilibrium}
S.~Deffner and E.~Lutz.
\newblock Nonequilibrium work distribution of a quantum harmonic oscillator.
\newblock {\em Physical Review E}, 77(2):021128, 2008.

\bibitem{blickle2006thermodynamics}
V.~Blickle, T.~Speck, L.~Helden, U.~Seifert, and C.~Bechinger.
\newblock Thermodynamics of a colloidal particle in a time-dependent
  nonharmonic potential.
\newblock {\em Physical review letters}, 96(7):070603, 2006.

\bibitem{harbola2006quantum}
U.~Harbola, M.~Esposito, and S.~Mukamel.
\newblock Quantum master equation for electron transport through quantum dots
  and single molecules.
\newblock {\em Physical Review B}, 74(23):235309, 2006.

\bibitem{esposito2006fluctuation}
M.~Esposito and S.~Mukamel.
\newblock Fluctuation theorems for quantum master equations.
\newblock {\em Physical Review E}, 73(4):046129, 2006.

\bibitem{kheirandish2020many}
F.~Kheirandish.
\newblock Many-body work distributions.
\newblock {\em Physics Letters A}, page 126296, 2020.

\bibitem{campisi2011erratum}
M.~Campisi, P.~H{\"a}nggi, and P.~Talkner.
\newblock Erratum: Colloquium: Quantum fluctuation relations: Foundations and
  applications [rev. mod. phys. 83, 771 (2011)].
\newblock {\em Reviews of Modern Physics}, 83(4):1653, 2011.

\bibitem{talkner2008microcanonical}
P.~Talkner, P.~H{\"a}nggi, and M.~Morillo.
\newblock Microcanonical quantum fluctuation theorems.
\newblock {\em Physical Review E}, 77(5):051131, 2008.

\bibitem{crooks1999entropy}
G.~E. Crooks.
\newblock Entropy production fluctuation theorem and the nonequilibrium work
  relation for free energy differences.
\newblock {\em Physical Review E}, 60(3):2721, 1999.

\bibitem{jarzynski2004nonequilibrium}
C.~Jarzynski.
\newblock Nonequilibrium work theorem for a system strongly coupled to a
  thermal environment.
\newblock {\em Journal of Statistical Mechanics: Theory and Experiment},
  2004(09):P09005, 2004.

\bibitem{jarzynski2011equalities}
C.~Jarzynski.
\newblock Equalities and inequalities: Irreversibility and the second law of
  thermodynamics at the nanoscale.
\newblock {\em Annu. Rev. Condens. Matter Phys.}, 2(1):329--351, 2011.

\bibitem{evans1993probability}
D.~J Evans, E.~Godert~D. Cohen, and G.~P. Morriss.
\newblock Probability of second law violations in shearing steady states.
\newblock {\em Physical review letters}, 71(15):2401, 1993.

\bibitem{talknerLutzHangi2007}
E.~Lutz P.~Talkner and P.~Hänggi.
\newblock Fluctuation theorems: Work is not an observable.
\newblock {\em Physical review E}, 75(5):050102(R), 2007.

\bibitem{EspositoHarbolaMukamel2009}
U.~Harbola M.~Esposito and S.~Mukamel.
\newblock Nonequilibrium fluctuations, fluctuation theorems, and counting
  statistics in quantum systems.
\newblock {\em Reviews of Modern Physics}, 81(4):1665, 2009.

\bibitem{cereceda2000quantum}
J.~L. Cereceda.
\newblock Quantum mechanical probabilities and general probabilistic
  constraints for einstein--podolsky--rosen--bohm experiments.
\newblock {\em Foundations of Physics Letters}, 13(5):427--442, 2000.

\bibitem{schrodinger1935gegenwartige}
E.~Schr{\"o}dinger.
\newblock Die gegenw{\"a}rtige situation in der quantenmechanik i, ii, iii,
  naturwiss 23, 807, 823, 844.
\newblock {\em English translation: Quantum Theory and Measurement (Trans:
  Wheeler, JA, Zurek, WH (eds.))(Princeton University Press, Princeton, 1983)},
  1935.

\bibitem{walls1983squeezed}
D.~F. Walls.
\newblock Squeezed states of light.
\newblock {\em nature}, 306(5939):141--146, 1983.

\bibitem{slusher1985observation}
R.~E. Slusher, L.~W. Hollberg, B.~Yurke, J.~C. Mertz, and J.~F. Valley.
\newblock Observation of squeezed states generated by four-wave mixing in an
  optical cavity.
\newblock {\em Physical Review Letters}, 55(22):2409, 1985.

\bibitem{vahlbruch2008}
H.~Vahlbruch, M.~Mehmet, S.~Chelkowski, B.~Hage, A.~Franzen, N.~Lastzka,
  S.~Gossler, K.~Danzmann, and R.~Schnabel.
\newblock Observation of squeezed light with 10-db quantum-noise reduction.
\newblock {\em Physical review letters}, 100(3):033602, 2008.

\bibitem{kimble1977photon}
H.~J. Kimble, M.~Dagenais, and L.~Mandel.
\newblock Photon antibunching in resonance fluorescence.
\newblock {\em Physical Review Letters}, 39(11):691, 1977.

\bibitem{gerber2009intensity}
S.~Gerber, D.~Rotter, L.~Slodi{\v{c}}ka, J.~Eschner, H.~J. Carmichael, and
  R.~Blatt.
\newblock Intensity-field correlation of single-atom resonance fluorescence.
\newblock {\em Physical review letters}, 102.

\bibitem{rempe1987observation}
G.~Rempe, H.~Walther, and N.~Klein.
\newblock Observation of quantum collapse and revival in a one-atom maser.
\newblock {\em Physical review letters}, 58(4):353, 1987.

\bibitem{lipfert2018time}
T.~Lipfert, F.~Krumm, M.~I Kolobov, and W.~Vogel.
\newblock Time ordering in the classically driven nonlinear jaynes-cummings
  model.
\newblock {\em Physical Review A}, 98(6):063817, 2018.

\bibitem{brune1992manipulation}
M.~Brune, S.~Haroche, J.~M. Raimond, L.~Davidovich, and N.~Zagury.
\newblock Manipulation of photons in a cavity by dispersive atom-field
  coupling: Quantum-nondemolition measurements and generation of
  ‘‘schr{\"o}dinger cat’’states.
\newblock {\em Physical Review A}, 45(7):5193, 1992.

\bibitem{slosser1989harmonic}
J.~J. Slosser, P.~Meystre, and S.~L. Braunstein.
\newblock Harmonic oscillator driven by a quantum current.
\newblock {\em Physical review letters}, 63(9):934, 1989.

\bibitem{guo1996generation}
G.-C. Guo and S.-B. Zheng.
\newblock Generation of schr{\"o}dinger cat states via the jaynes-cummings
  model with large detuning.
\newblock {\em Physics Letters A}, 223(5):332--336, 1996.

\bibitem{weidinger1999trapping}
M.~Weidinger, B.~TH Varcoe, R.~Heerlein, and H.~Walther.
\newblock Trapping states in the micromaser.
\newblock {\em Physical Review Letters}, 82(19):3795, 1999.

\bibitem{brattke2001generation}
S.~Brattke, B.~T. Varcoe, and H.~Walther.
\newblock Generation of photon number states on demand via cavity quantum
  electrodynamics.
\newblock {\em Physical review letters}, 86(16):3534, 2001.

\bibitem{rabi1936process}
I.~I. Rabi.
\newblock On the process of space quantization.
\newblock {\em Physical Review}, 49(4):324, 1936.

\bibitem{esteve2004quantum}
D.~Esteve, J.-M. Raimond, and J.~Dalibard.
\newblock {\em Quantum entanglement and information processing: lecture notes
  of the Les Houches Summer School 2003}.
\newblock Elsevier, 2004.

\bibitem{wallraff2004strong}
A.~Wallraff, D.~I. Schuster, A.~Blais, L.~Frunzio, R.-S. Huang, J.~Majer,
  S.~Kumar, S.~M. Girvin, and R.~J. Schoelkopf.
\newblock Strong coupling of a single photon to a superconducting qubit using
  circuit quantum electrodynamics.
\newblock {\em Nature}, 431(7005):162--167, 2004.

\bibitem{chiorescu2004coherent}
I.~Chiorescu, P.~Bertet, K.~Semba, Y.~Nakamura, C.~J. P.~M. Harmans, and J.~E.
  Mooij.
\newblock Coherent dynamics of a flux qubit coupled to a harmonic oscillator.
\newblock {\em Nature}, 431(7005):159--162, 2004.

\bibitem{hatakenaka1996josephson}
N.~Hatakenaka and S.~Kurihara.
\newblock Josephson cascade micromaser.
\newblock {\em Physical Review A}, 54(2):1729, 1996.

\bibitem{sornborger2004superconducting}
A.~T Sornborger, A.~N Cleland, and M.~R. Geller.
\newblock Superconducting phase qubit coupled to a nanomechanical resonator:
  Beyond the rotating-wave approximation.
\newblock {\em Physical Review A}, 70(5):052315, 2004.

\bibitem{meier2004spin}
F.~Meier and D.~D. Awschalom.
\newblock Spin-photon dynamics of quantum dots in two-mode cavities.
\newblock {\em Physical Review B}, 70(20):205329, 2004.

\bibitem{basset2013single}
J.~Basset, D.-D. Jarausch, A.~Stockklauser, T.~Frey, C.~Reichl, W.~Wegscheider,
  T.~Markus Ihn, K.~Ensslin, and A.~Wallraff.
\newblock Single-electron double quantum dot dipole-coupled to a single
  photonic mode.
\newblock {\em Physical Review B}, 88(12):125312, 2013.

\bibitem{kasprzak2010up}
J.~Kasprzak, S.~Reitzenstein, E.~A. Muljarov, C.~Kistner, C.~Schneider,
  M.~Strauss, S.~H{\"o}fling, A.~Forchel, and W.~Langbein.
\newblock Up on the jaynes--cummings ladder of a quantum-dot/microcavity
  system.
\newblock {\em Nature materials}, 9(4):304--308, 2010.

\bibitem{Freitas2017Josephson}
N.~Hatakenaka and S.~Kurihara.
\newblock Josephson cascade micromaser.
\newblock {\em Physical Review A}, 54(2):1729, 1996.

\bibitem{jarzynski2007comparison}
C.~Jarzynski.
\newblock Comparison of far-from-equilibrium work relations.
\newblock {\em Comptes Rendus Physique}, 8(5-6):495--506, 2007.

\bibitem{jarzynski1997nonequilibrium}
C.~Jarzynski.
\newblock Nonequilibrium equality for free energy differences.
\newblock {\em Physical Review Letters}, 78(14):2690, 1997.

\bibitem{plastina2014irreversible}
F.~Plastina, A.~Alecce, T.~J.~G Apollaro, G.~Falcone, G.~Francica, F.~Galve,
  N.~Lo. Gullo, and R.~Zambrini.
\newblock Irreversible work and inner friction in quantum thermodynamic
  processes.
\newblock {\em Physical review letters}, 113(26):260601, 2014.

\end{thebibliography}
\bibliographystyle{unsrt}

\end{document}